\documentclass[11pt]{article}
\newcommand{\sect}[1]{\setcounter{equation}{0}\section{#1}}

\def\spazio#1{\vrule height#1em width0em depth#1em}

\def\fraz#1#2{{\strut\displaystyle #1\over\displaystyle #2}}
\def\esp#1{e^{\displaystyle#1}}

\def\L{{\cal{L}}}

\def\HE{{\cal{H}}_E}
\def\P{{\cal{P}}}
\def\xp{\dot x}

\begin{document}

\title{\bf Spinning particle in an
external linearized gravitational wave field}

\author{A. Barducci and R.Giachetti}

\date{}

\maketitle \centerline{{  Department of Physics, University of
Florence  and I.N.F.N. Sezione di Firenze }}\centerline{{ Via G.
Sansone 1, I-50019 Sesto Fiorentino , Firenze, Italy }}
\bigskip

\centerline{{ Firenze Preprint - DFF - 419/07/04}}

\bigskip

\noindent

\begin{abstract}
We study the interaction of a scalar and a spinning particle with
a coherent linearized gravitational wave field treated as a
classical spin two external field. The spin degrees of freedom of
the spinning particle are described by skew-commuting variables.
We derive the explicit expressions for the eigenfunctions and the
Green's functions of the theory. The discussion is exact within the
approximation of neglecting radiative corrections and we prove
that the result is completely determined by the semiclassical
contribution.
\end{abstract}
\bigskip

\sect{Introduction}

For many years Grassmann variables have been a common practice
for giving a {\it pseudoclassical} framework to describe spin,
\cite{GO}-\cite{GRR}, or other internal degrees of freedom of
elementary particles, \cite{bcl2}. Of course the value of such
models must be thought of in view of their quantization, that can be
the canonical one or, in view of the inherent pseudoclassical
description, the path-integral quantization. Although some
technical caution had to be taken in extending the
path-integral to skew-commuting variables (it is, for instance,
relevant whether their number is even or odd), nevertheless the
program was successfully carried out, see {\it e.g.} \cite{BBC}.

We have recently presented the quantization of the theory
describing a spin $1\over 2$  particle in an external electromagnetic wave field
by using the path-integral and external source method and we have
produced the corresponding Feynman propagator, \cite{BGJP}. The
electromagnetic interaction at the pseudoclassical level was
introduced by the standard minimal coupling directly in the
constraints describing the theory. The self-consistency of the
theory automatically produced the non minimal electromagnetic
coupling characteristic for the spin 1/2 particle (Pauli term).
The calculations leading to the final result were straightforward,
although somewhat complex, due to the above mentioned cautions. In
this paper we consider the problem of describing the interaction
of a scalar and a spinning particle at a first quantized level
with an external gravitational field. Of course, if we don't want
to go far beyond our possibilities of performing explicit
calculations, the gravitational field must be taken in the linear
approximation. Even with this simplifying assumption, however, we
have to face new features with respect to \cite{BGJP}, such as the
treatment of covariant derivatives of Grassmann variables,
responsible for the spin-gravitational field interaction and whose
affine connections have to be substituted with their symmetric
part, implying that the spinning particle cannot be directly
coupled to the torsion, \cite{bcl3}. Following the approach
started by \cite{volkovlandau} we are able to evaluate the
eigenfunctions and the Feynman kernel  if we assume that the
linearized field is that of a gravitational wave. When restricting
ourselves to the even sector, we recover the propagator of a scalar
particle, calculated in \cite{barduccigiachetti3}, by using
symmetry arguments and canonical transformations in the
path-integral framework.

A brief summary of the paper is as follows. In Section 2 we give
the main definitions of the physical quantities as well as the
gauge conditions we will assume on the metric and on the affine
connections. We then calculate the wave function for the scalar
particle, whose Green's function is discussed in Section 3
together with its semiclassical limit. The linearized theory for a
spinning particle in the external gravitational wave is described
in Section 4, while in the final Section 5 we consider the
corresponding Green's function and we show, for the spinning
particles too, the semiclassical nature of the results.

\sect{Scalar particle in an external gravitational wave:
linear theory}

According to general relativity, the interaction of
matter with a given gravitational field is introduced in a
geometrical way by replacing the flat Minkowski metric
$\eta_{\mu\nu}=(+,-,-,-)$ by a
tensor $g_{\mu\nu}(x)$ depending upon
the coordinates
$x^\mu=(x^0,\overrightarrow{x})\equiv(t,\overrightarrow{x})$
and by using the general covariance principle
\cite{weinberg}. Therefore the action describing a massive scalar
particle interacting with an external gravitational field can be
written as:
\begin{equation}
S=\int{d^4 x}\,\sqrt{-g(x)}\, \L(\phi(x),D_{\mu}\phi(x))
\label{scalaraction}
\end{equation}
where $g(x)$ is the determinant of the metric and $D_\mu$
the covariant derivative. 
Letting  $g^{\mu\nu}(x)$ be the inverse of the metric tensor, the
Lagrangian density $\L$ has the form
\begin{equation}
\L={1\over2}\,g^{\mu\nu}(x)\, D_ {\mu}\phi(x)\,
D_{\nu}\phi(x) -{1\over2} m^2 \phi^2(x)
\label{scalarlagrangian}
\end{equation}
and produces the equation of motion
\begin{equation}
\Bigl(\,g^{\mu\nu}(x)\,D_{\mu}\,D_{\nu}+m^2
\,\Bigr)\,\phi(x)=0\,.
\label{equationmotion}
\end{equation}
The energy-momentum tensor of the matter field,in the absence of
gravity, is obtained by the usual relation
\begin{equation}
T_0^{\mu\nu}(x)=-{2\over\sqrt{-g(x)}} \,{\delta S \over\delta
g_{\mu\nu}(x)}
\Biggl|_{g_{\mu\nu}(x)=\eta_{\mu\nu}}=-\eta^{\mu\nu} \L_0 +
\partial^{\mu}\phi(x) \,\partial^{\nu}\phi(x)
\label{energymomentumtensor}
\end{equation}
where $\L_0$ is the free Lagrangian density.
When the gravitational field is weak and described by a
small perturbation to the flat metric
\begin{equation}
g_{\mu\nu}(x)\simeq \eta_{\mu\nu}
+h_{\mu\nu}(x),~~~~~~~~~~g^{\mu\nu}(x)\simeq \eta^{\mu\nu}
-h^{\mu\nu}(x)
\label{weakfield}
\end{equation}
we can consider the linearized theory with a
Lagrangian density
\begin{equation}
\L=\L_0-{1\over2} h_{\mu\nu}(x) ~T_0^{\mu\nu}(x)=
 (1+{1\over2}h(x))\L_0-{1\over2}h^{\mu\nu}(x)\,\partial_{\mu}\phi(x)
\,\partial_{\nu}\phi(x)
\label{linearizedlagr}
\end{equation}
where $h(x)=\eta_{\mu\nu}h^{\mu\nu}(x)=h^\mu_\mu(x) $. The Lagrange
equations derived from  (\ref{linearizedlagr}) are thus
\begin{equation}
\Bigl(\,(1+{1\over2}h(x)) \,(\partial^{\mu}\partial_{\mu}
+m^2)-h^{\mu\nu}(x)\,
\partial_{\mu}\partial_{\nu}-(\partial_{\mu}h^\mu_\nu(x)-
{1\over2}\partial_{\nu}h(x))\,\partial^{\nu}\,\Bigr)\,\phi=0
\label{linearizedequation}
\end{equation}
Upon multiplying on the left by the factor $~(1-h(x)/2)~$,
at this order in $h_{\mu\nu}(x)$ equation (\ref{linearizedequation})
can be further simplified to
\begin{equation}
\Bigl(\,\partial^{\mu}\partial_{\mu}-h^{\mu\nu}(x)\,
\partial_{\mu}\partial_{\nu}-(\partial_{\mu}h^\mu_\nu(x)-
{1\over2}\partial_{\nu}h(x))\,\partial^{\nu}+m^2\,\Bigr)\phi=0
\label{linearizedequationbis}
\end{equation}
From (\ref{linearizedequationbis}) it appears that we are
considering a theory describing the interaction of scalar matter
with a given gravitational field treated as a spin two external
field in a flat Minkowski spacetime \cite {borde}.

As is well known there are some conditions that can
be imposed on the metric tensor, in analogy to the gauge choice for the
vector potential $A_\mu(x)$ in electrodynamics. A particularly convenient
requirement  is represented by the harmonic condition \cite
{weinberg}
\begin{equation}
g^{\mu\nu}(x)\, \Gamma{^\lambda_{\mu\nu}}(x)=0
\label{harmonic}
\end{equation}
whose linearization reads
\begin{equation}
\partial_{\mu}h^\mu_\nu(x) = {1\over2}\partial_{\nu} h(x)
\label{linearizedharmonic}
\end{equation}
To our purpose it turns out to be even more convenient to choose a
restriction of the general harmonic gauge, that is
\begin{equation}
\partial_{\mu}h^\mu_\nu(x) = 0, \qquad h(x)=0
\label{linearizedharmonicbis}
\end{equation}
This choice, in addition to some algebraic simplifications,
also avoids the more serious ordering problem that would
be present in some terms of
equation (\ref{linearizedequation}). Finally we consider
the linearized gravitational field as the one produced by a
wave of arbitrary spectral composition and
polarization properties, but  propagating in a fixed direction:
\begin{equation}
g_{\mu\nu}(x)\simeq \eta_{\mu\nu} +h_{\mu\nu}(\kappa_x),
\qquad h_{\mu\nu}(\kappa_x)=a_ {\mu\nu}\, f(\kappa_x)\,, \qquad
\kappa_x=k\!\cdot\! x\,.
\label{kweaklinearized}
\end{equation}
Here $k^2=0$ and $a_ {\mu\nu}=a_ {\nu\mu}$ is the polarization
tensor. Due to these choices the equation
(\ref{linearizedharmonicbis})  then becomes 
\begin{equation}
k_\mu a^\mu_\nu = 0, ~~~~~~~ a{^\mu_\mu}=0
\label{klinearizedharmonicbis}
\end{equation}
and  it is easy to cast the linearized Klein-Gordon wave
equation (\ref{linearizedequation}) into the form
\begin{equation}
(\,\partial^{\mu}\partial_{\mu}-h^{\mu\nu}(x)\,\partial_{\mu}\partial_{\nu}
+m^2\,)\,\phi(x)=0
\label{finalinearizedequation}
\end{equation}

Remembering the derivation of the Volkov solution for the problem
of spin $1\over 2$ particle in a electromagnetic wave field
\cite{BGJP}, \cite{volkovlandau}, we look for a solution of
(\ref{finalinearizedequation}) in the form
\begin{equation}
\phi_p(x)=(2p_0\,V)^{-{1\over 2}}\,
\exp\Bigl(\,-i p\!\cdot\! x-iF(\kappa_x)\,\Bigr),
\label{trialsolution}
\end{equation}
where $p^\mu$ is a constant 4-vector and $F(\kappa_x)\rightarrow 0$
as  $x^0\rightarrow-\infty$.
Moreover, since $F(\kappa_x)$ is still an undetermined function,
by adding to
$p^\mu$ an arbitrary 4-vector proportional to $k^\mu$ we obtain a wave
function $\phi_p(x)$ of the same form. Therefore, without loss of
generality we can impose on the 4-vector $p^\mu$ the mass-shell
condition $p^2 = m^2$.
After substituting the solution (\ref{trialsolution}) into equation
(\ref{finalinearizedequation}) and taking into account the
mass-shell relation, the null character of $k^\mu$ and the
harmonic gauge condition (\ref{linearizedharmonicbis}) we obtain
for $F(\kappa_x)$ the following first-order differential equation:
\begin{equation}
2\,k\!\cdot\! p~ F\,' (\kappa_x)-p_\mu\, p_\nu \,h^{\mu\nu}(\kappa_x)=0
\label{fscalarequation}
\end{equation}
where $F\,'(\kappa_x)$ is the derivative with respect to the
argument. By integrating this first order equation, from
(\ref{fscalarequation}) we finally obtain for $\phi_p(x)$ the
Volkov-like solution
\begin{equation}
\phi_p(x)=
(2p_0\,V)^{-{1\over 2}}\,
\exp\,\Bigl(-i p\!\cdot\! x-{i\over2\,k\!\cdot\!
p}\int\limits_{-\infty}^{\kappa_x}{d\kappa}~~
h^{\mu\nu}{(\kappa)}~p_\mu ~p_\nu\,\Bigr)\,.
\label{volkovsolution}
\end{equation}

\sect{Green's function for the scalar particle}

The standard way of obtaining the Green's function for the wave
equation   is to solve the non
homogeneous boundary value problem
with an added delta-function source term
(see {\it e.g.} \cite{schwinger}):
\begin{equation}
(\,\partial^{\mu}\partial_{\mu}-h^{\mu\nu}(x)\,
\partial_{\mu}\partial_{\nu}
+m^2\,)\,{\Delta_F}(x,y)= - {\delta^4}(x-y)\,.
\label{greenscalarequation}
\end{equation}
By extracting the singularity due to the mass shell condition, we can
assume for the Green function the form
\begin{equation}
\Delta_F(x,y)=\int {d^{4}p \over (2 \pi ) ^{4}} {\esp{-i p\!\cdot\!
(x-y)}\over p^2-m^2+i\epsilon} {\esp{-iF_1(\kappa_x,\kappa_y)}}
\label{greenscalaransatz}
\end{equation}
where $\kappa_x = k\!\cdot\! x$, $\kappa_y = k\!\cdot\! y$.
The complete Feynman's propagator
${\Delta_F}(x,y)$ is therefore a superposition of free propagators
modulated by the phase factor $\exp\,\{-iF_1(\kappa_x,\kappa_y)\}$.
The equation for
$F_1(\kappa_x,\kappa_y)$ is determined by
(\ref{greenscalarequation}-\ref{greenscalaransatz}) and reads
\begin{eqnarray}
{}&& \int {d^{4}p \over (2 \pi ) ^{4}}\, {\esp{-i p\!\cdot\!
(x-y)-iF_1(\kappa_x,\kappa_y)}\over p^2-m^2+i\epsilon}
 {\,\Bigl[\, 2\, k\!\cdot\! p \,{\partial F_1(\kappa_x,\kappa_y)\over\partial\
{\kappa_x}} - p_\mu\, p_\nu\, h^{\mu\nu} (\kappa_x)\,\Bigr]} =
\phantom{XXXX}\spazio{1.4}\cr
{}&&\phantom{XXXXXXXXXXXXXXXX}\int {d^{4}p \over (2 \pi ) ^{4}}\,
 {\esp{-i p\!\cdot\! (x-y)}
\,\Bigl(\,1- \esp{-iF_1(\kappa_x,\kappa_y)}}\,\Bigr)
\label{f1equation}
\end{eqnarray}

The term in square brackets on the
left-hand side of (\ref{f1equation})  is just the equation
 (\ref{fscalarequation}). Therefore the left-hand
side of (\ref{f1equation}) vanishes if we take
\begin{equation}
F_1(\kappa_x,\kappa_y) = {1\over 2 \,k\!\cdot\! p}
\int\limits_{\kappa_y}^{\kappa_x} {d \kappa}\,\, p_\mu\, p_\nu\,
h^{\mu\nu} (\kappa) \label{f1solution}
\end{equation}
We now verify that the integral on the right-hand side of
(\ref{f1equation}) vanishes
too. We first separate the component of $p_\mu$ in
the direction of $k_\mu$ \cite{kibble},
$~p^\mu = {p\,'\,}^\mu + \alpha\, k^\mu \label{pandk}~$
where $p'^{\,\mu}$ spans a three-dimensional surface. Then the
function $F_1(\kappa_x,\kappa_y)$ is independent of $\alpha$ and
we can integrate over $\alpha$ getting
\begin{eqnarray}
{}&& \int{d\alpha\over 2\pi} \,{\esp{-i \alpha (\kappa_x-\kappa_y)}
\Bigl(\, 1- \esp{-iF_1(\kappa_x,\kappa_y)}}\,\Bigr) =
\delta(\kappa_x-\kappa_y)\,\Bigl(\,1-
{\esp{-iF_1(\kappa_x,\kappa_y)}}\,\Bigr)\,.\phantom{XXX}
\label{verificaf1}
\end{eqnarray}
The right hand side of (\ref{verificaf1}) is obviously
vanishing in view
of (\ref{f1solution}). The Green's function has therefore the form
\begin{equation}
\Delta_F(x,y)=\int {d^{4}p \over (2 \pi ) ^{4}}\, {\esp{-i p\!\cdot\!
(x-y)}\over p^2-m^2+i\epsilon}\,\,
\exp\,\Bigl(-{i\over2\,k\!\cdot\!
p}\int\limits_{\kappa_y}^{\kappa_x}{d\kappa}~
h^{\mu\nu}{(\kappa)}~p_\mu ~p_\nu\,\Bigr)
\label{deltafsolution}
\end{equation}
This Feynman kernel was derived in paper \cite{barduccigiachetti3}
by looking at the problem in a semiclassical way and by evaluating
the quantum mechanical path-integral in a closed form.

We would now like to comment on the classical nature of the results so
far obtained for the wave function (\ref{volkovsolution}) and the
Green's function (\ref{deltafsolution}). Indeed the classical action
for a relativistic scalar particle in an external
gravitational field,
\begin{equation}
S= -m \int\limits_{\tau_i}^{\tau_f}\Bigl(\,g_{\mu\nu}(x)\, \xp^\mu
\,\xp^\nu\,\Bigr)^{1/2} \,{d\tau}
\label{actionparticle}
\end{equation}
is obtained by integrating a singular Lagrangian, giving rise to
the first class constraint $\chi=\,g^{\mu\nu}(x)\,
p_{\mu}\,p_{\nu}- m^2~$ and to a vanishing canonical Hamiltonian.
An extended Hamiltonian proportional to the constraint,
$\HE(x,p)=\alpha_1 \,\chi~$  is then defined. For a linearized
gravitational wave field as (\ref{kweaklinearized}) and in the
harmonic gauge (\ref{linearizedharmonicbis}) we have
\begin{equation}
\HE(x,p)=\alpha_1 (\,\eta^{\mu\nu}\, p_{\mu}\,p_{\nu}-
h^{\mu\nu}(x)\, p_{\mu}\,p_{\nu}- m^2\,)
\label{extendedhamiltonianlin}
\end{equation}
Writing the corresponding covariant Hamilton-Jacobi equation
\begin{equation}
\eta^{\mu\nu}\,\, {\partial S(x,p)\over \partial x^\mu}\,
{\partial S(x,p)\over
\partial x^\nu} - h^{\mu\nu}(x)\,\, {\partial S(x,p)\over
\partial x^\mu}\, {\partial S(x,p)\over \partial x^\nu} - m^2 = 0
\label{hjequation}
\end{equation}
we shall look for a solution  of the characteristic Hamilton
function $S(x,p)$, expressed as a function of the coordinates
$x^\mu$ and the initial momentum $p^\mu$, of the
general form
\begin{equation}
S(x,p) = -p\!\cdot\! x -S_1(\kappa_x)\,. \label{hjgeneralsolution}
\end{equation}
From (\ref{hjequation}) and (\ref{klinearizedharmonicbis})
we get the differential equation
\begin{equation}
2\, k\!\cdot\! p\,\, S'_1(\kappa_x) - h^{\mu\nu}(\kappa_x)\,\,
p_\mu\, p_\nu =0\,, \label{hjequationbis}
\end{equation}
identical to equation (\ref{fscalarequation}) for $F(\kappa_x)$.
Hence
\begin{equation}
S(x,p) =
-p\!\cdot\! x - {1\over 2\, k\!\cdot\! p} \int\limits_{-\infty
}^{k\cdot x} {d \kappa}\,\, h^{\mu\nu}(\kappa)\,\,
p_\mu\, p_\nu \label{hjsolution}
\end{equation}
so that the Volkov-like solution (\ref{volkovsolution}) turns out
to be
\begin{equation}
\phi_p(x) = (2 p_0 V)^{-{1\over 2}}\,\, \esp{i S(x,p)}
\label{hjvolkovsolution}
\end{equation}
The WKB method applied to
(\ref{finalinearizedequation}) therefore gives exact result
in the first order approximation, {\it i.e.} we have found an
exact classical solution of equation
(\ref{finalinearizedequation}).

\sect{Dirac particle in an external gravitational wave:
linear theory}

We now study the same problem for a spin $1\over2$ particle, considering
the gravitational wave as an external field defined
in a flat Minkowski background. (For the
connection with the canonical intrinsic approach using Dirac
equation in curved space-time, see \cite{borde} ). Consequently
the minimal linearized
Lagrangian describing the interaction  of fermionic matter with a
given gravitational field treated as a spin two external field in
a Minkowski background is given by the expression
\cite{weinberg,borde}
\begin{equation}
\L=\L_0-{1\over2} h_{\mu\nu}(x) ~T_0^{\mu\nu}(x)\,,
\label{fermioniclagrangian}
\end{equation}
in complete analogy with the scalar case described by eq.
(\ref{linearizedlagr}). The Lagrangian density of the free
fermionic matter $\L_0$  and the corresponding
symmetrized energy-momentum tensor
$T_0^{\mu\nu}(x)$ are given by the equations

\begin{equation}
\L_0 = {1\over 2}\, \bar \psi(x)\,(\,i\gamma^\mu
\overrightarrow{\partial_\mu}-m\,)\,\psi(x) +{1\over 2}\,
\bar \psi(x)\,(\,-i\gamma^\mu
\overleftarrow{\partial_\mu}-m\,)\,\psi(x)
\label{freefermioniclagrangian}
\end{equation}
\begin{equation}
T_0^{\mu\nu}(x)=-\eta^{\mu\nu} \L_0
+{i\over 4}\,
 \Bigl(\,\bar \psi(x)\, (\,\gamma^\mu
\overrightarrow{\partial^{\,\nu}}- \gamma^\mu
\overleftarrow{\partial^{\,\nu}}\,)\,\psi(x)
+ \bar \psi(x)\, (\,\gamma^\nu
\overrightarrow{\partial^{\,\mu}}- \gamma^\nu
\overleftarrow{\partial^{\,\mu}}\,)\,\psi(x)\,\Bigr)
\label{fermionictensor}
\end{equation}

\noindent The Lagrangian density $\L$ then becomes
\begin{equation}
\L= (1+{1\over2}h(x))\,\L_0-{i\over4}\, h^{\mu\nu}(x)\,
\bar \psi(x)\,(\,\gamma_\mu
\overrightarrow{\partial_\nu}- \gamma_\mu
\overleftarrow{\partial_\nu}\,)\,\psi(x)
\label{fermioniclagrangianbis}
\end{equation}
and it can also be considered as obtained from the
Lagrangian density established in general relativity for the
interaction
of the Dirac field with a prescribed gravitational field in the
linear approximation \cite{weinberg},\cite{borde}.
The  Dirac equation we derive
from (\ref{fermioniclagrangianbis}) turns out to be \cite{borde}
\begin{equation}
\Bigl(\,(1+{1\over2}h(x))\, (\,i \gamma^{\mu} \partial_{\mu} - m\,)-
{i\over2}\,
h^{\mu\nu}(x)\, \gamma_\mu {\partial_\nu} -
{i\over4}\,\partial{_\nu} h{^{\mu\nu}}(x)\gamma_\mu +
{i\over4}\,\partial{_\mu} h(x)\gamma^\mu\,\Bigr)\,\psi(x)=0
\label{diracequation}
\end{equation}
or, multiplying it again by $(1-{1\over2}h(x))$ and taking the linear
terms in
 $h_{\mu\nu}(x)$,
\begin{equation}
\Bigl(\,(\,i \gamma^{\mu} \partial_{\mu} - m\,)-{i\over2}\,
h^{\mu\nu}(x)\,
\gamma^\mu {\partial_\nu} -
{i\over4}\,\partial{_\nu} h{^{\mu\nu}}(x)\gamma_\mu +
{i\over4}\,\partial{_\mu} h(x)\gamma^\mu\,\Bigr)\,\psi(x)=0
\label{diracequationbis}
\end{equation}
The harmonic condition (\ref{linearizedharmonicbis}) further simplifies 
the wave equation (\ref{diracequationbis}) giving the final form
\begin{equation}
(\,i \gamma^{\mu} \partial_{\mu}-{i\over2}\, h^{\mu\nu}(x)\,
\gamma_\mu {\partial_\nu} -m\,)\,\psi(x)=0
\label{diracequationfinal}
\end{equation}
Moreover, since the external field represents a linearized
gravitational wave, we assume again for
$h_{\mu\nu}(x)$ the expression $h_{\mu\nu}(\kappa_x)$ given in
(\ref{kweaklinearized}-\ref{klinearizedharmonicbis}).

Instead of solving the first order Dirac equation
(\ref{diracequationfinal}), in analogy to what is done
for the electromagnetic interactions \cite{volkovlandau},
 we find it easier to
consider the second order equation obtained by applying to
(\ref{diracequationfinal}) the operator
$(\,i\gamma^\mu\,\partial_\mu - {i\over2} h^{\mu\nu}(\kappa_x)\gamma_\mu
\partial_\nu+m\,)\,$. By using the
supplementary gauge condition (\ref{linearizedharmonicbis}) we
finally get
\begin{equation}
\Bigl(\,\partial^{\mu}\partial_{\mu}-h^{\mu\nu}(\kappa_x)
\partial_{\mu}\partial_{\nu}+{i\over2}
{\sigma_{\mu\nu}}\, (\partial^{\mu}h^{\nu\rho}(\kappa_x))\, \partial_{\rho}
+m^2\,\Bigr)\,\psi(x)=0
\label{quadraticdiracequation}
\end{equation}
If we compare (\ref{quadraticdiracequation})
with the scalar analogous
equation (\ref{finalinearizedequation}), we can see
that they are identical but for a term representing the
spin-gravitational field interaction, as should be expected.
Hence we shall assume for the solutions of
(\ref{quadraticdiracequation}) a general form
\begin{equation}
\psi_{p,s}(x)=(2 p_0 V)^{-{1\over 2}}\,\,\esp{-i p\!\cdot\!
x}\,G(\kappa_x)~u(p,s) \label{diracansatz}
\end{equation}
with the initial condition
\begin{equation}
\psi_{p,s}(x)
{{}\atop
{\longrightarrow\atop {x^0\, \rightarrow\,-\infty}}}
(\,2 p_0 V\,)^{-{1\over 2}}\,
\esp{-i p\!\cdot\! x}\,\, {u(p,s)}
\label{diracboundary}
\end{equation}
where $u(p,s)$ is a constant spinor which is a solution of the
corresponding free Dirac equation
$~(\,\gamma^{\mu} p_\mu - m\,)\, u(p,s)=0\,.\,$
Using the relations
$k^2 = k_\mu\, h^{\mu\nu}(\kappa_x)=h_{\mu}^\mu(\kappa_x)=0$
and $p^2= m^2$,
from (\ref{diracansatz}) and
(\ref{quadraticdiracequation}) we get the first order
differential equation
\begin{equation}
2\,k\!\cdot\! p\,~ G\,' (\kappa_x)+
i \Bigl(\,h^{\mu\nu}(\kappa_x)\, p_\mu p_\nu
-{1\over2} \, k^\mu p_\nu\,
{dh^{\nu\rho}(\kappa_x)\over d\kappa_x}\,{\sigma_{\rho\mu}}\,
\Bigr)\,G(\kappa_x) =0
\label{gequation}
\end{equation}
with solution
\begin{equation}
G(\kappa_x)=\exp\Bigl({-{i\over2\,k\!\cdot\!
p}\int\limits_{-\infty}^{\kappa_x}{d\kappa}~
(\,h^{\mu\nu}{(\kappa)}\,p_\mu \,p_\nu
- {1\over2}
\,
k^\mu p_\nu \,{dh^{\nu\rho}(\kappa)\over d\kappa}\,{\sigma_{\rho\mu}}\,)
}\,\,\Bigr)\,.
\label{ansatzsolution}
\end{equation}
Taking into account that
$(\, k^\mu p_\nu\, a^{\nu\rho}\,\sigma_{\rho\mu}\,)^n = 0$
for $n>1$, after some straightforward simplifications
the Volkov-like solution
(\ref{diracansatz}) can finally be put into the form
\begin{equation}
{\psi(x)}=
(2 p_0 V)^{-{1\over 2}}
\Bigl(
1 + {i\over4}
\, k^\mu p_\nu\, {h^{\nu\rho}(\kappa_x){\sigma_{\rho\mu}}\,\Bigr)\,
{\exp\Bigl({{-i p\!\cdot\! x} -{i\over2\,k\!\cdot\!
p}\int\limits_{-\infty}^{\kappa_x}{d\kappa}\,\,
h^{\mu\nu}{(\kappa)}\,p_\mu p_\nu }}}\Bigr)\, u(p,s)
\label{volkovdiracsolution}
\end{equation}
where, as usual, $\sigma_{\mu\nu} = {i\over2}[\gamma_\mu,\gamma_\nu]$
\cite{bjorkendrell} .

\sect{Green's function for the Dirac particle}

We present here a direct calculation of the electron Green's
function in the presence of an external gravitational plane wave.
The Green's function satisfies the equation
\begin{equation}
(\,i \gamma^{\mu} \partial_{\mu}-{i\over2} h^{\mu\nu}(\kappa_x)
\gamma_\mu
{\partial_\nu}-m \,)\, S_F (x,y)= {\delta^4}(x-y)
 \label{fermionicgreen}
\end{equation}
with the usual associated boundary conditions.
As in Section 4 we prefer
working with a second order equation. In fact, if we let
\begin{equation}
S_F (x,y)= (\,i \gamma^{\mu} \partial_{\mu}-{i\over2} h^{\mu\nu}
(\kappa_x)
\gamma_\mu {\partial_\nu}+m \,)\,{\Delta_F}(x,y)
\label{fermionicbosonic}
\end{equation}
then, using again the  gauge condition (\ref{linearizedharmonicbis}),
we find that ${\Delta_F}(x,y)$  satisfies
\begin{equation}
(\,\partial^{\mu}\partial_{\mu}-h^{\mu\nu}(\kappa_x)
\,\partial_{\mu}\partial_{\nu}
+{i\over2}\,
{\sigma_{\mu\nu}}\,(\partial^{\,\mu}h^{\nu\rho}(\kappa_x))\,
\partial_{\rho}
+m^2\,)\,\Delta_F (x,y)= -{\delta^4}(x-y)
\label{fermionicquadratic}
\end{equation}
Equation (\ref{fermionicquadratic}) is identical to
the analogous equation (\ref{greenscalarequation}) for the scalar
particles but for the spin term $~(i/2)\,\sigma_{\mu\nu}\,
(\partial^{\,\mu} h^{\nu\rho}(\kappa_x))\,\partial_\rho\,$ .
It is therefore
natural to look for a solution of the form
\begin{equation}
\Delta_F(x,y)=\int {d^{4}p \over (2 \pi ) ^{4}}\, {\esp{-i p\!\cdot\!
(x-y)}\over p^2-m^2+i\epsilon}
~{\esp{-iF_1(\kappa_x,\kappa_y)}}~N(\kappa_x, \kappa_y)
\label{ansatzfermionicquadratic}
\end{equation}
where $F_1(\kappa_x,\kappa_y)$ is the same as in the
scalar case (see eq. (\ref{f1solution})) and $N(\kappa_x,\kappa_y)$
is an unknown function to be determined. From
(\ref{ansatzfermionicquadratic}) and (\ref{fermionicquadratic})
the equation for $N(\kappa_x,\kappa_y)$ becomes
\begin{eqnarray}
{}&&
\int {d^{4}p \over (2 \pi ) ^{4}} {\esp{-i p\!\cdot\!
(x-y)-iF_1(\kappa_x,\kappa_y)}\over p^2-m^2+i\epsilon}\,
\Bigl(
\,{1\over2} \,\, k^\mu p_\nu\,
{dh^{\nu\rho}(\kappa_x)\over d\kappa_x}\,{\sigma_{\rho\mu}}\,
N (\kappa_x,\kappa_y)\,+
\phantom{XXXXXX}\spazio{1.4}\cr
{}&& \,2\, i k\,\!\cdot\! p
{\partial N(\kappa_x,\kappa_y)\over \partial\
{\kappa_x}} \,\Bigr)
= \int {d^{4}p \over (2 \pi ) ^{4}} \esp{-i p\!\cdot\! (x-y)} \,\Bigl(\,
1- N(\kappa_x,\kappa_y)\, {\esp{-i F_1(\kappa_x,\kappa_y)}}\,
\Bigr)\phantom{XXX}
\label{ansatzfermionicequation}
\end{eqnarray}
The left-hand side of eq. (\ref{ansatzfermionicequation}) vanishes
by choosing
\begin{equation}
N(\kappa_x,\kappa_y) =  \exp\,\Bigl(\,{i\over4\,k\!\cdot\! p}\,
\int\limits_{\kappa_y}^{\kappa_x}{d\kappa}~
\, k^\mu\, p_\nu\,
{dh^{\nu\rho}(\kappa)\over d\kappa}\,{\sigma_{\rho\mu}}\,
\Bigr)
\label{ansatzfermionicsolution}
\end{equation}
and, as in the scalar case, we
easily verify that the right-hand side vanishes too.
Thus
\begin{eqnarray}
{}&& \Delta_F(x,y) = \int {d^{4}p \over (2 \pi ) ^{4}}\,\Bigl(
\,1 +
{i\over4\,k\!\cdot\! p}\,
\, k^\mu p_\nu\,
{h^{\nu\rho}(\kappa_x)}\,{\sigma_{\rho\mu}}\,
\,\Bigr)\,
\esp{{-i p\!\cdot\! x} -{i\over2k\!\cdot\!p}
\int\limits_{-\infty}^{k\cdot x}{d\kappa}~
h^{\mu\nu}{(\kappa)}\,p_\mu p_\nu}\cr
{}&& {1 \over{ p^2-m^2+i\epsilon}}~\,\Bigl(
\,1 -
{i\over4\,k\!\cdot\! p}\,
\, k^\mu p_\nu\,
{h^{\nu\rho}(\kappa_y)}\,{\sigma_{\rho\mu}}\,
\,\Bigr)\, \esp{{i p\!\cdot\! y}
+{i\over 2\,k\!\cdot\!
p}\int\limits_{-\infty}^{k\cdot y}{d\kappa}~
h^{\mu\nu}{(\kappa)}\,p_\mu p_\nu}
\label{fermionicquadraticsolution}
\end{eqnarray}
is a solution of the inhomogeneous second order Dirac equation
(\ref{fermionicquadratic}) and from equation
(\ref{fermionicbosonic}) we finally get for the Feynman propagator
${S_F}(x,y)$ the expression
\begin{eqnarray}
{}&& S_F (x,y)= \Bigl(\,i \gamma^{\mu}
\partial_{\mu}-{i\over2}\, h^{\mu\nu}(\kappa_x)
\gamma_\mu {\partial_\nu}+m
\,\Bigr)\,\Delta_F(x,y)
%\phantom{XXXXXXXXXXXXXXXXXX}
\spazio{1.4}\cr
{}&&
%\phantom{S_F (x,y)}
=\int {d^{4}p \over (2 \pi ) ^{4}}
\,
\Bigl(
\,1 +
{i\over4\,k\!\cdot\! p}\,
\, k^\mu p_\nu\,
{h^{\nu\rho}(\kappa_x)}{\sigma_{\rho\mu}}
\,\Bigr)\,
\,
{{\gamma_\mu
p^\mu+m}\over p^2-m^2+i\epsilon} \,
\Bigl(
\,1 -
{i\over4\,k\!\cdot\! p}
\, k^\mu p_\nu\,
{h^{\nu\rho}(\kappa_y)}{\sigma_{\rho\mu}}
\,\Bigr)\,
\spazio{1.4}\cr
{}&&
\phantom{XXXXXXXXX}
\exp\Bigl(\,
{{-i p\!\cdot\! (x-y)} -{i\over2\,k\!\cdot\! p}\int\limits_{\kappa_y}
^{\kappa_x}{d\kappa}~ h^{\mu\nu}{(\kappa)}\,p_\mu p_\nu}
\,\Bigr)
\label{fermionicpropagator}
\end{eqnarray}

As we have already discussed for the scalar particle in Sect.3, it
would be interesting to show the $\it pseudoclassical$ nature of the
results obtained for the Dirac particle too. We start with
the pseudoclassical Lagrangian for
a spin-${1\over 2}$ particle in an
arbitrary external gravitational field \cite{bcl3}
\begin{equation}
\L=-\fraz i2 g_{\mu\nu}(x)\, \zeta^\mu \,({\dot \zeta}^\nu +
\Gamma{^\nu_{\lambda\rho}}(x) {\dot x}^\rho \zeta^\lambda) - \fraz
i2\zeta_5{\dot \zeta}_5- m ~
\Bigl(\,g_{\mu\nu}(x)\,({\dot x}^\mu-\fraz
im \zeta^\mu{\dot\zeta}_5)\,({\dot x}^\nu-\fraz im
\zeta^\nu{\dot\zeta}_5)\,\Bigr)^{1\over 2}
\label{pseudoclassicalaction}
\end{equation}
where $\Gamma{^\nu_{\lambda\rho}}(x)$ are the usual Christoffel
symbols and the  Grassmann variables $\zeta^\mu$ are quite
naturally introduced, since in quantization they correspond to the
use of the spin matrices conform to metric $\gamma^\mu$ (see for
instance \cite{wheeler} and references therein.) By introducing
the $\it vierbein$ field $G^A_\mu(x)$ and its inverse $H^\mu_B(x)$
\begin{eqnarray}
&{}& g_{\mu\nu}(x)=\eta_{AB}\,G{^A_\mu}(x)\, G{^B_\nu}(x)\,,
\qquad ~ g^{\mu\nu}(x)=\eta^{AB}(x)\,H{^\mu_A}(x)\, H{^\nu_B}(x)
\label{ghvierbein}\\
\spazio{1.6}
&{}& {H^\mu_A}(x)\, {G^A_\nu}(x)  = {\delta^\mu_\nu}\,,\qquad \qquad
\qquad
{H^\mu_A}(x)\, {G^B_\mu}(x)  = {\delta^B_A}
\label{inversevierbein}
\end{eqnarray}
we can transform a world four-vector $\zeta^\mu$ into a
``local'' one and we can also introduce Grassmann variables $\xi^A$
corresponding, after quantization, to the usual Dirac matrices
(flat space spin matrices)
\begin{equation}
\xi^A = G{^A_\mu}(x)\,\zeta^\mu\,,
\qquad \zeta^\mu = H{^\mu_A}(x)\,\xi^A\,,
\qquad\xi_5 = \zeta_5\,.
\label{worldlocal}
\end{equation}

The singular Lagrangian (\ref{pseudoclassicalaction}) gives rise
to the two first class constraints $~\chi_D = \P_\mu \,\zeta^\mu -
i m \,(\pi_5-(i/2)~\zeta_5)~$ and $~\chi =g^{\mu\nu}(x)\,\P_\mu
\P_\nu -m^2~$, where $P_\mu$ is the canonical momentum and $\P_\mu
= P_\mu -(i/2) \,g_{\rho\nu}(x)\,\zeta^\rho\,
\Gamma{^\nu_{\mu\lambda}}(x)\,\zeta^\lambda$  is the mechanical
one. We also have a second class constraint
\begin{equation}
\pi_\mu = \fraz i2\,g_{\mu\nu}(x)\,\zeta^\nu
\label{secondconstraint}
\end{equation}
and, in addition, we can further require \cite{bcl1,bcl3}
\begin{equation}
\pi_5=-\fraz i2\,\zeta_5\,. \label{constraint5}
\end{equation}
The relations (\ref{secondconstraint}-\ref{constraint5}) form a
set of second-class constraints whose relevant Dirac brackets
are
\begin{equation}
\{\zeta_5,\zeta_5\} =-i\,,\qquad
\{\zeta^\mu,\zeta^\nu \}=ig^{\mu\nu}(x)\,, \qquad\{\xi^A,\xi^B\} =i{\eta^{AB}}. \label{diracbrackets}
\end{equation}

Some remarks on the quantization are now in order. The Dirac brackets
(\ref{diracbrackets}) give the anticommutation rules
\begin{equation}
[\,{\widehat\xi}_5,{\widehat\xi}_5 \,]_+= 1
\,,\qquad
[\,{\widehat\xi}^A,{\widehat\xi}^B \,]_+= -\eta^{AB}\,
,\qquad
[\,{\widehat\zeta}^\mu,{\widehat\zeta}^\nu \,]_+= -g^{\mu\nu}(x)\,.
\label{anticommutation}
\end{equation}
The first and the second relation can be satisfied by putting
\begin{equation}
 {\widehat\xi}_5=
2^{\,-{1\over 2}}\,\gamma_5 \,,\qquad
{\widehat\xi}^A= 2^{\,-{1\over 2}}\,\gamma_5\,\gamma^A\,\label{gammalocal}
\end{equation}
where $\gamma_5$ and $\gamma^A$ are the usual flat space Dirac
matrices. We must now choose a representation for the algebra of
the operators ${\widehat\zeta}^\mu$. A possible choice which will
correspond to the use of the spin matrices conform to metric is
\begin{equation}
\widehat\zeta^\mu=H{^\mu_A}(x)\,\widehat\xi^A =
2^{\,-{1\over 2}}\,\gamma_5\,\gamma^\mu \label{gammaworld}
\end{equation}
where
\begin{equation}
[\gamma^\mu,\gamma^\nu ]_+= 2 g^{\mu\nu}(x) \label{diracalgebra}
\end{equation}
As usual, the extended Hamiltonian $\HE$ is
written as a linear combination of the first-class constraints
with arbitrary coefficients and the simplest choice appears to be
\begin{equation}
\HE=\alpha_1~ (g^{\mu\nu}(x)\, \P_\mu \P_\nu- m^2)\,.
\end{equation}
The constraint
$\chi_D$ becomes $\chi_D = \P_\mu \,\zeta^\mu - m\,\zeta_5$, once
eq. (\ref{constraint5}) is accounted for.
It could be used, after quantization, for constructing the spinorial
states. Considering again the restriction to a linearized
gravitational wave field in the harmonic gauge,
(\ref{kweaklinearized}-\ref{klinearizedharmonicbis})
and taking into account the linearized form
$H{^\mu_A}(\kappa_x)= \delta {^\mu_A} - (1/2)\,  h{^\mu_A}(\kappa_x)$,
$G{^A_\mu}(\kappa_x)= \delta {^A_\mu} + (1/2)\,  h{^A_\mu}(\kappa_x)$
of the vierbein fields
we finally get
\begin{equation}
\HE=\alpha_1\, (\eta^{\mu\nu}\, P_\mu P_\nu -
h^{\mu\nu}(\kappa_x)\, P_\mu P_\nu + i
P_\rho {\partial h^{\rho\nu}(\kappa_x)\over \partial x_ {\mu}}\,
\xi_\mu\,
\xi_\nu - m^2) \label{linearizeddirachamiltonian}
\end{equation}
Therefore the variables $\zeta^\mu$ satisfy the linearized relations
\begin{equation}
\zeta^\mu = \xi^\mu -{1\over 2} h{^\mu_A}(x)\, \xi^A
\label{zetalimit}
\end{equation}
and we finally obtain the following equation for the Hamilton's
characteristic function $S(x)$:
\begin{equation}
\eta^{\mu\nu}\, {\partial S(x)\over {\partial x^\mu}}
{\partial S(x)\over
\partial x^\nu} - h^{\mu\nu}(x)\, {\partial S(x)\over
{\partial x^\mu}} {\partial S(x)\over {\partial x^\nu}} -
i\, {\partial S(x)\over {\partial x^\rho}}
{\partial h^{\rho\nu}(x)\over {\partial x_\mu}}\xi_\mu \xi_\nu
- m^2 =0\,.
\label{dirachamiltonjacobi}
\end{equation}

The solution  (\ref{hjsolution}) obtained in
the scalar case suggests, for the pseudoclassical Hamilton
characteristic function, the ansatz
\begin{equation}
S(x)=  -p\!\cdot\! x - {1\over 2\, k\!\cdot\! p} \int\limits_{-\infty
}^{\kappa_x} {d \kappa}~p_\mu p_\nu \,h^{\mu\nu}(\kappa) +
S_D(\kappa_x)\,. 
\label{fermionichjansatz}
\end{equation}
The unknown function $S_D$ will satisfy the first-order differential
equation
\begin{equation}
2\, {k\!\cdot\! p}\, {dS_D(\kappa_x)\over d\kappa_x} + i k^\mu p_\nu
{dh^{\nu\rho}(\kappa_x)\over d\varphi_x}\,   \xi_\rho \xi_\mu =0
\label{sdequation}
\end{equation}
with solution
\begin{equation}
S_D(x)=  - {i\over 2\, k\!\cdot\! p} \int\limits_{-\infty
}^{\kappa_x} {d \kappa} ~
k^\mu p_\nu {dh^{\nu\rho}(\kappa)\over d\kappa} \,\xi_\rho\, \xi_\mu
\label{fermionichjsolution}
\end{equation}

It can easily be seen that the operator ${\widehat S}$
obtained upon quantization  -- through the use of eqs. (\ref{gammalocal})
-- of the pseudoclassical $S(x)$ given in
(\ref{fermionichjansatz})
can be represented as
\begin{equation}
{\widehat S}=  -p\!\cdot\! x - {1\over 2 \,k\!\cdot\! p}
\int\limits_{-\infty }^{\kappa_x} {d \kappa} ~\Bigl(\,h^{\mu\nu}
(\kappa)\, p_\mu p_\nu -{1\over 2} k^\mu p_\nu {dh^{\nu\rho}(\kappa)
\over
d\kappa}\, {\sigma_{\rho\mu}}\,\Bigr) \label{fermionichjsolutionoperator}
\end{equation}
Therefore by applying the operator
$~(2 p_0 V)^{-{1\over 2}}\,\esp {i \widehat S}~$ to a free
constant spinor $u(p,s)$ we see that the Volkov-like solution
(\ref{volkovdiracsolution}) will be written as follows
\begin{equation}
\psi(x) = (2 p_0 V)^{-{1\over 2}}\,{\esp{i\widehat S}}~
u(p,s)
\label{finalequation}
\end{equation}
completely showing the semiclassical nature of the solution.

\end{document}